\documentclass[twocolumn,superscriptaddress,
amsmath,amssymb,aps,prl]{revtex4-1}

\usepackage{graphicx}
\usepackage{physics}
\usepackage{comment}
\usepackage{color}
\usepackage{dcolumn}
\usepackage{bm}

\usepackage[colorlinks,urlcolor=blue,citecolor=blue,linkcolor=blue]{hyperref}

\renewcommand{\k}{{\bf k}}

\newcommand{\q}{{\bf q}}
\newcommand{\Q}{{\bf Q}}

\newcommand{\ef}{E_F}

\newcommand{\nn}{\nonumber}
\newcommand{\beq}{\begin{equation}}
\newcommand{\eeq}{\end{equation}}

\begin{document}

\title{Exact quantum virial expansion for the optical response of doped two-dimensional semiconductors}

\author{Brendan C. Mulkerin}
\thanks{B. C. M. and A. T. contributed equally to this work.}
\affiliation{School of Physics and Astronomy, Monash University, Victoria 3800, Australia}
\affiliation{ARC Centre of Excellence in Future Low-Energy Electronics Technologies, Monash University, Victoria 3800, Australia}

\author{Antonio Tiene}
\thanks{B. C. M. and A. T. contributed equally to this work.}
\affiliation{Departamento de F\'isica Te\'orica de la Materia
  Condensada \& Condensed Matter Physics Center (IFIMAC), Universidad
  Aut\'onoma de Madrid, Madrid 28049, Spain}
\affiliation{School of Physics and Astronomy, Monash University, Victoria 3800, Australia}

\author{Francesca~Maria~Marchetti}
\affiliation{Departamento de F\'isica Te\'orica de la Materia
  Condensada \& Condensed Matter Physics Center (IFIMAC), Universidad
  Aut\'onoma de Madrid, Madrid 28049, Spain}

\author{Meera M. Parish}
\affiliation{School of Physics and Astronomy, Monash University, Victoria 3800, Australia}
\affiliation{ARC Centre of Excellence in Future Low-Energy Electronics Technologies, Monash University, Victoria 3800, Australia}

\author{Jesper Levinsen}
\affiliation{School of Physics and Astronomy, Monash University, Victoria 3800, Australia}
\affiliation{ARC Centre of Excellence in Future Low-Energy Electronics Technologies, Monash University, Victoria 3800, Australia}

\date{\today}

\begin{abstract}
We introduce a quantum virial expansion for the optical response of a doped two-dimensional semiconductor. As we show, this constitutes a perturbatively exact theory in the high-temperature or low-doping regime, where the electrons' thermal wavelength is smaller than their interparticle spacing.
We obtain exact analytic expressions for the photoluminescence and we predict new features such as a non-trivial shape of the attractive branch peak  related to universal resonant exciton-electron scattering and an associated energy shift from the trion energy. 
Our theory furthermore allows us to formally unify the two distinct theoretical pictures that have been applied to this system, where we reveal that the predictions of the conventional trion picture correspond to a  high-temperature and weak-interaction limit of Fermi-polaron theory.
Our results are in excellent agreement with recent experiments on doped monolayer MoSe$_2$ and they provide the foundation for modelling a range of emerging optically active materials such as van der Waals heterostructures.
\end{abstract}

\maketitle

The problem of a quantum impurity interacting with a background medium represents a paradigmatic example of a strongly correlated many-body system. First considered in the context of electrons moving in a crystal lattice~\cite{Landau1933}, the quantum impurity (polaron) problem has since been generalized to many different systems across a wide range of energy scales, including magnetic impurities~\cite{Kondo1964}, quantum mixtures in ultracold atomic gases~\cite{Massignan_Zaccanti_Bruun,Levinsen2Dreview,Scazza2022}, and even protons in neutron stars~\cite{kutscheraProtonImpurityNeutron1993}. Most recently, it has provided key insights into the optical response of doped two-dimensional (2D) semiconductors~\cite{Suris2003correlation,Ross2014,Sidler_NatPhys_2017,Efimkin2017,Rana_PRB2020,Rana_PRB2021,Efimkin_PRB_2021,Goldstein_JCP2020,Xiao_JPCL2021,Liu_NatComm2021,Zipfel_PRB2022,Huang2023}. Here, an optically excited exciton (bound electron-hole pair) is immersed in a fermionic medium of charge carriers (electrons or holes), leading to the appearance of two peaks in the optical response --- a ``repulsive polaron'' and an ``attractive polaron'' --- at energies which evolve, respectively, into those of the exciton and a trion (charged exciton bound state) in the limit of vanishing doping. 
Such exciton polarons have attracted a large amount of interest since they can be realized in atomically thin transition metal dichalcogenides (TMDs), where there is the prospect of technological applications involving both charge doping and coupling to light~\cite{Mak_NatPh2016,Wang2018}. 

However, there is an ongoing debate about whether the optical response should be described within a Fermi-polaron picture~\cite{Suris2003correlation,Sidler_NatPhys_2017,Efimkin2017,Rana_PRB2020,Rana_PRB2021,Efimkin_PRB_2021,Katsch_PRB2022}, where excitons are coherently dressed by excitations of the medium to form new polaronic quasiparticles~\cite{chevy2006}, or whether it is better to use a more conventional trion picture~\cite{Lampert_PRL58,Deilmann_PRL2016,Tempelaar-Berkelbach_NatCommr2019} which involves independent few-body states (excitons and trions). The two pictures have been shown to give indistinguishable results for some observables (e.g., the oscillator strength) at low charge doping~\cite{GlazovJCP2020}, but this requires an intrinsic linewidth that exceeds the Fermi energy of the charge carriers\footnote{See footnote 60 in Ref.~\cite{GlazovJCP2020}}.
On the other hand, in the context of ultracold atoms, it is known that at zero temperature there is a density-driven transition between a Fermi polaron and few-body bound states equivalent to trions~\cite{Parish_PRA11,Parish_PRA2013,Vlietinck2014,Kroiss2014,Levinsen2Dreview,qu2022efficient}. 
Importantly, the nature of the exciton polaron has implications for other properties such as the transport of optical excitations under an electric field~\cite{Cotlet_PRX2019} and optical nonlinearities~\cite{Tan2020,Kyriienko_PRL2020,Bastarrachea-Magnani_PRL2021}.

\begin{figure}
    \centering
    \includegraphics[width=.95\columnwidth]{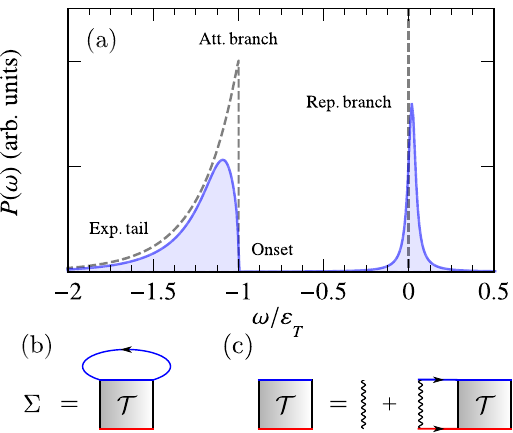}\centering
    \caption{(a) Schematic illustration (solid [blue] lines) of the key features of PL from a doped semiconductor at low doping and/or high temperature where $E_F\ll T$. For the attractive branch, this includes the exponential tail related to electron recoil, the shape of the onset due to resonant %
    electron-exciton scattering, and the shift of the peak from the trion energy. Here, we have neglected any additional exciton broadening due to effects beyond those described in the Hamiltonian \eqref{eq:ham}, such as disorder and radiative recombination. The (gray) dashed lines are the predictions from the conventional trion theory~\cite{Esser_PRB2000,Zipfel_PRB2022}. (b) Leading-order contribution to the exciton self-energy within the virial expansion, describing the interaction between the exciton (red line) and an electron (blue line). (c) Diagrams contributing to the two-body $T$ matrix (square) due to the exciton-electron interaction potential (wavy line).
    }
    \label{fig:diagrams-schematic}
\end{figure}

In this Letter, we resolve this question and reveal that these two pictures are in fact connected when we account for the crucial role played by temperature. 
We introduce a quantum virial expansion~\cite{Liu2013} for the optical response,
which we show is \emph{perturbatively exact} when the temperature $T$ greatly exceeds the 
Fermi energy $E_F$, and is therefore applicable at high temperature and/or low doping.
We show that this corresponds to a limit of the Fermi-polaron picture where the coherent dressing cloud of the attractive polaron quasiparticle is destroyed by thermal fluctuations (see Ref.~\cite{finiteTlong} for details), 
in contrast to the situation at lower temperatures.
We demonstrate that the virial expansion predicts hitherto unrecognized features in photoluminescence (PL)
such as  
a non-trivial behavior of the attractive peak near the trion energy 
related to 2D resonant exciton-electron scattering, and a Lorentzian repulsive peak,
as illustrated in Fig.~\ref{fig:diagrams-schematic}(a). We compare our results to recent experiments on doped MoSe$_2$ monolayers~\cite{Zipfel_PRB2022} and find excellent agreement, which implies that the trion binding energy has been previously overestimated. 
Finally, we show analytically that the virial expansion reduces to the predictions of the trion picture in the limit where $\ef\to0$.

\textit{Model}.---We model a doped 2D semiconductor 
using the following Hamiltonian for excitons and excess charge carriers:
\begin{align} \label{eq:ham}
    \hat{H}= & 
    \sum_{\k} \! \left(\epsilon_{\k}\hat{c}^{\dag}_{\k}\hat{c}^{}_{\k}+\epsilon_{X\k}\hat{x}^{\dag}_{\k}\hat{x}^{}_{\k} 
    \right) -
    \sum_{\k\k'\q} \! v_\q \, \hat{x}^{\dag}_{\k}\hat{c}^{\dag}_{\k'}\hat{c}^{}_{\k'+\q}\hat{x}^{}_{\k-\q} \;.
\end{align}
Since the optically generated exciton is tightly bound, we treat it as a structureless boson, 
with corresponding operator $\hat{x}^{\dag}_{\k}$, mass $m_X$ and free-particle dispersion $\epsilon_{X\k}=|\k|^2/2m_X \equiv
k^2/2m_X$, where the energy is measured from that of the 1$s$ exciton at rest. The fermionic operator $\hat{c}^{\dag}_{\k}$ creates charge carriers (electrons or holes) with mass $m$ and dispersion $\epsilon_{\k}=k^2/2m$. For simplicity, we generally assume the charge carriers are electrons, but note that our results equally hold for the hole-doped case. 
We also ignore the spin/valley degree of freedom and 
consider spin-polarized electrons that are 
distinguishable from the electron 
within the exciton, since this is sufficient to describe the polaron and trion physics in TMDs such as MoSe$_2$ monolayers~\cite{Sidler_NatPhys_2017,Huang2023}. 
Here, and in the following, we set $\hbar = k_B= 1$ and work in a system of unit area.

The second term in Eq.~\eqref{eq:ham} describes the 
attractive charge-(induced) dipole interactions between electrons 
and excitons, which give rise to a trion bound state~\cite{Fey_PRB_2020,Efimkin_PRB_2021}. Note that we can treat the trion as an effective two-body (electron-exciton) bound state since the exciton binding energy exceeds the trion binding energy $\varepsilon_T$ by an order of magnitude in TMDs~\cite{Wang2018}. 
Furthermore, the potential $v_\q$ is sufficiently short ranged that it can be described with a low-energy $s$-wave scattering amplitude~\cite{landau2013quantum}, scaling as $1/r^4$ at large exciton-electron separation $r$.
We neglect the interactions between 
electrons 
since these are not necessary to describe the trion bound state and 
they do not contribute to the leading order behavior in the high-temperature limit, $T/\ef \gg 1$, as we discuss below. To be specific, we assume that the temperature and density are such that the electrons form a Fermi liquid rather than a strongly correlated Wigner crystal~\cite{Platzman1974}. At the same time, we assume that $T$ is sufficiently low such that phonon scattering effects can be safely neglected.

While we formulate our results in the language of excitons in a Fermi sea of charge carriers such as electrons, our results apply more generally to any dilute gas of impurities interacting via short-range interactions with a 
2D Boltzmann gas. In particular, the features of the spectrum discussed below would also be observable in cold-atom experiments on 2D Fermi gases~\cite{Koschorreck2012,Zhang2012,Oppong2019}, and our theory can straightforwardly be extended to the three-dimensional case.

\textit{Photoluminescence}.---
The starting point of our analysis is the detailed balance relation~\cite{Kennard_18,vanRoosbroeck_54,Kubo_57,Martin-Schwinger_59,Stepanov_57} between optical absorption, which is proportional to the exciton spectral function $A(\omega)$, and photoluminescence $P(\omega)$,
\begin{align}
    P(\omega)&=e^{-\beta\omega}A(\omega)\;,
\label{eq:PL_func}
\end{align}
with $\beta\equiv 1/T$. This expression is valid (up to an unimportant frequency-independent prefactor~\cite{finiteTlong}) within linear response for a system in thermal equilibrium, under the assumption of a low density of excitons such that they can effectively be treated as uncorrelated. 

The spectral function is related to the (retarded) exciton Green's function $G_X$ via $A(\omega)=-\frac{1}{\pi}\Im G_X(\omega+i0)$, where the factor $+i0$ signifies that here and in the following the poles are shifted slightly into the lower half of the complex $\omega$ plane~\cite{fetterbook} (at this stage, we do not explicitly introduce the exciton linewidth which can mask the intrinsic features of the PL). The Green's function in turn satisfies the Dyson equation
\begin{align}\label{eq:dyson}
    G_X(\omega)=\frac1{\omega-\Sigma(\omega)}\;,
\end{align}
in terms of the self-energy $\Sigma(\omega)$ for the zero-momentum exciton. Calculating the PL thus amounts to obtaining the self-energy which, in the general case of a strongly correlated system, can only be done approximately.

\textit{Virial expansion}.---A key insight is that we can apply the quantum virial expansion to the exciton self-energy at finite temperature. Specifically, this corresponds to a systematic
expansion in powers of the fugacity $z=e^{\beta\mu}$, where $\mu$ is the chemical potential of the background Fermi gas.
In the high-temperature/low-doping regime $T\gtrsim E_F$, we have $z \lesssim1$, allowing us to perform an exact perturbative expansion around the ideal Boltzmann gas limit of the medium (where $z \simeq \beta E_F$). 
The virial expansion has been extensively used in other contexts. For instance, to obtain thermodynamic quantities and quantum corrections to the equation of state in condensed matter physics~\cite{Vedenov1959,Siddon1974}, nuclear physics~\cite{Horowitz2006} and %
ultracold gases~\cite{Ho2003,Liu2013}. It has also been used to calculate response functions for atomic gases~\cite{Hu2010,Schaefer2D,ngamp13,Barth2014,Ngamp15,sun2015,Sun2017}, magnetic impurities~\cite{Larkin1971}, magnons~\cite{Nishida2013} and Coulomb systems~\cite{Hofmann_2013}.

At lowest order in $z$, only two-point correlations involving the exciton and electrons are present in the self-energy, corresponding to all the ladder diagrams depicted in Fig.~\ref{fig:diagrams-schematic}(b,c). Crucially, higher $N$-point correlations with $N >2$ only enter at higher order in $z$ since they require multiple electrons to be scattered from the medium, where each medium excitation is weighted by $z$~\cite{Leyronas2011,sun2015}. This furthermore means that we can neglect electron-electron interactions if we work at lowest order in $z$, where the medium corresponds to a Boltzmann gas. Thus, the leading order exciton self-energy takes the form
\beq
    \Sigma(\omega)= z \sum_{\q} e^{-\beta \epsilon_{\q}} \mathcal{T}(\q_r;\q,\omega+\epsilon_\q) \, .
\label{eq:Xself-energy}
\eeq
Here the $T$ matrix $\mathcal{T}(\k;\Q,\omega)$ describes the sum of repeated scattering processes between an exciton and an electron in vacuum, where $\Q$ and $\omega$ are the total momentum and energy, respectively, while $\k$ is the electron-exciton relative momentum (where the incoming and outgoing momenta are equal). Note that, due to Galilean invariance, the center-of-mass and relative contributions 
separate. For a zero-momentum exciton with energy $\omega$ and an electron with kinetic energy $\epsilon_{\q}$, as in Eq.~\eqref{eq:Xself-energy}, the center-of-mass momentum is simply $\q$ and the relative momentum $\q_r = \q m_X/m_T$, with $m_T = m+m_X$.

In principle, one can obtain the self-energy \eqref{eq:Xself-energy} for an arbitrary electron-exciton 
$T$ matrix. 
However, since the relevant energy scales in TMDs (i.e., $T$, $E_F$, and the trion binding energy $\varepsilon_T$) are much smaller than that set by the range of $v_\q$, i.e., the exciton binding energy, 
the $T$ matrix is well approximated by its low-energy $s$-wave form~\cite{AdhikariAJP86}
\begin{align}
  \mathcal{T}(\q_r;\q,\omega) \simeq \mathcal{T}_0(\q, \omega) &= \frac{2\pi}{m_r} %
  \frac1{\ln [-\varepsilon_T/(\omega - \epsilon_{T\q})]},
   \label{eq:2b_Tmatrix}
\end{align}
with the reduced electron-exciton  mass $m_r=m_X m/m_T$ and the center-of-mass (trion) dispersion $\epsilon_{T\q}=q^2/2m_T$.
This is independent of the relative momentum and coincides with the limit of a zero-range potential, which has been shown to provide a good approximation for the interactions~\cite{Efimkin_PRB_2021,Huang2023}.
Together, Eqs.~(\ref{eq:PL_func}-\ref{eq:Xself-energy}) and \eqref{eq:2b_Tmatrix} allow us to straightforwardly calculate the optical response in the virial expansion.

The expressions in Eqs.~\eqref{eq:Xself-energy} and \eqref{eq:2b_Tmatrix} 
are in fact equivalent to the celebrated Chevy ansatz~\cite{chevy2006} for the Fermi polaron when we consider its finite-temperature generalization~\cite{Liu2019} and take the limit $z \ll 1$
(see also the accompanying paper~\cite{finiteTlong} for details). 
Thus, our approach is continuously connected to the Fermi-polaron picture of excitons in doped semiconductors which is based on the zero-temperature Chevy ansatz~\cite{Sidler_NatPhys_2017,Efimkin2017,Efimkin_PRB_2021}.

\textit{Features of the photoluminescence spectrum}.---The resulting exciton spectral function contains two well-separated peaks, as illustrated in Fig.~\ref{fig:diagrams-schematic}(a): a repulsive branch centered close to the exciton at $\omega=0$, and an attractive branch that is peaked for frequencies $\omega\lesssim-\varepsilon_T$. %
For the repulsive branch, the leading order in $z$ is obtained by taking $\omega=0$ in the self-energy. In this case we find, to logarithmic accuracy in $\beta\varepsilon_T$~\footnote{This assumes that the typical kinetic energy in Eq.~\eqref{eq:2b_Tmatrix} is set by the temperature, with logarithmic corrections neglected. The inclusion of $\gamma_{\rm E}$ is found by expanding the special function called $\nu(x)$ around the limit $T\to\infty$ and utilizing the Laplace transform of the $\sin(x)$ function~\cite{gradshteyn2014table}.},
\begin{align}\label{eq:SErep}
 \Sigma_{\rm rep}(0)\simeq %
 \frac{E_F(m/m_r)}{\pi^2+\ln^2(e^{\gamma_{\rm E}}\beta\varepsilon_T)}\left[\ln(e^{\gamma_{\rm E}}\beta \varepsilon_T)-i\pi\right]\; ,
\end{align}
with $\gamma_{\rm E} \simeq 0.5772$ the Euler-Mascheroni constant. 

In the regime $T \lesssim \varepsilon_T$, which is the situation in most current experiments, 
the dominant contribution to the attractive branch arises from the pole of the $T$ matrix when $\omega=-\varepsilon_T+\epsilon_{T\q}$, related to the trion bound state. Expanding Eq.~\eqref{eq:2b_Tmatrix} around the pole gives 
\begin{align}
    \label{eq:2b_Tmatrixpole}
    \mathcal{T}_{0}(\mathbf{q},\omega+i0) \simeq 
    \frac{Z_T}{\omega-\epsilon_{T\q}+\varepsilon_T+i0}\; ,
\end{align}
with $Z_T=\frac{2\pi\varepsilon_T}{m_r}$ the residue at the pole.
This allows us to obtain the self-energy by straightforward contour integration, with the result 
\begin{align}\label{eq:SEatt}
    \Sigma_{\rm att}(\omega) & \simeq -z \varepsilon_T ( \tfrac{m_T}{m_X} )^2 e^{\frac{m_T}{m_X} \beta  (\omega + \varepsilon_T)} \nn \\ &\hspace{1mm}\times
    \,\left[{\rm Ei}[- \tfrac{m_T}{m_X} \beta(\omega + \varepsilon_T)]+i\pi \Theta(-\omega-\varepsilon_T)\right]\;,
\end{align}
where $ {\rm Ei}\left( x \right)$ is the exponential integral. Note that the pole expansion of the $T$ matrix is exact for %
the imaginary part of $ \Sigma_{\rm att}$ but only approximate for the real part, with the latter becoming exact close to the trion energy.

Combining these results yields the spectral function $A(\omega)$ and, from Eq.~\eqref{eq:PL_func}, the PL 
\begin{align}\label{eq:PLtotal}
    P(\omega)\simeq-\frac1\pi e^{-\beta\omega}\Im \frac{\Theta(-\omega-\varepsilon_T)}{\omega-\Sigma_{\rm att}(\omega)}-\frac1\pi \Im\frac1{\omega-\Sigma_{\rm rep}(0)}\;,
\end{align}
in terms of the self-energies in Eqs.~\eqref{eq:SErep} and \eqref{eq:SEatt}.
Equation~\eqref{eq:PLtotal} is a key result of this work. We see that the repulsive branch is a Lorentzian peak at $\omega = \Re\Sigma_{\rm rep}(0)$ with width $\Gamma_{R}=\pi(m/m_r) \ef/[\pi^2+\ln^2(e^{\gamma_{\rm E}}\beta\varepsilon_T)]$, where both the width and position scale with $E_F$, similar to Fermi polaron theories~\cite{Ngampruetikorn2012,Schmidt_PRA2012,Efimkin2017}. However, for the attractive branch, we find that we cannot satisfy the condition $\omega = \Re\Sigma_{\rm att}(\omega)$, indicating that there is no attractive polaron quasiparticle in the limit $z\ll 1$, unlike for the quantum degenerate case $z>1$~\cite{finiteTlong}.
Instead, we have an asymmetric continuum  of trion states, %
with a sharp onset at $\omega=-\varepsilon_T$ %
and an exponential tail involving trions and recoil electrons at finite relative momentum, where $P(\omega)\propto e^{\beta \omega m/m_X}/\omega^2$ for $-\omega\gg\varepsilon_T$ in agreement with Ref.~\cite{Esser_PRB2000}. Moreover, in the limit of an infinitely heavy exciton, we see that the tail in PL loses its exponential dependence, becoming a power law, unlike in the case of absorption.  
The shape of the onset is dictated by 2D resonant electron-exciton scattering at the trion energy, 
leading to a universal logarithmic divergence in the self-energy:
$\Sigma_{\rm att}(\omega\lesssim-\varepsilon_T)\simeq -z \varepsilon_T ( \tfrac{m_T}{m_X} )^2 \left [
\ln\left( -e^{\gamma_{\rm E}}\beta \tfrac{m_T}{m_X}(\omega+\varepsilon_T) \right) +i\pi \right]$. 
Previous trion theories of PL~\cite{Esser_PRB2000,Esser2001,Zipfel_PRB2022} focussed on the imaginary part of the self-energy, as we show below, and thus appear to have missed this divergence in the real part.

\begin{figure}
    \centering
    \includegraphics[width=0.82\columnwidth]{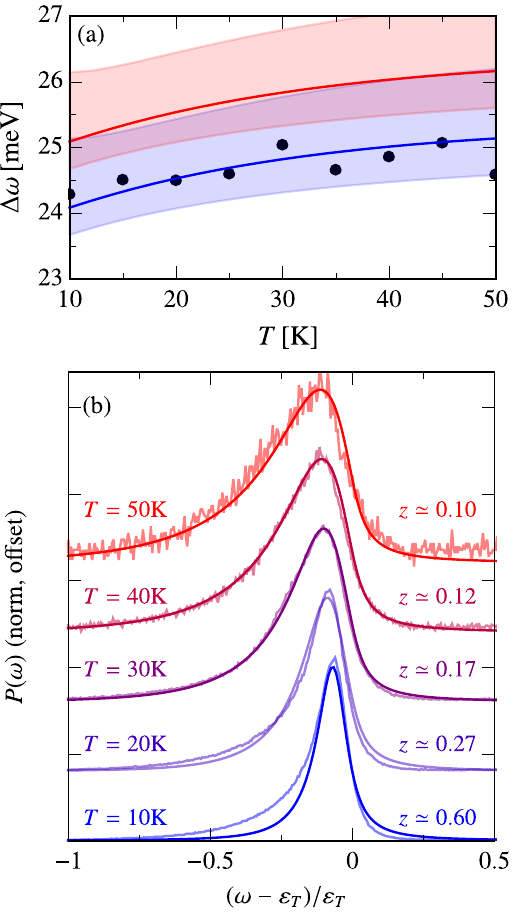}
    \caption{Photoluminescence in a hole-doped MoSe$_2$ monolayer. (a) Frequency difference between attractive and repulsive peaks as a function of temperature. The black 
    dot symbols are the experimental peak positions  obtained from Ref.~\cite{Zipfel_PRB2022}. The blue and red shaded regions correspond to the results of the virial expansion using binding energies $\varepsilon_T=22.5$~meV and $23.5$~meV, respectively. The solid lines correspond to the experimental hole density $n_h=0.5\times10^{11}$~cm$^{-2}$, and the lower and upper bounds of each shaded region to densities of $0.25\times10^{11}$~cm$^{-2}$  and $10^{11}$~cm$^{-2}$, respectively. (b) Comparison between theoretical (solid dark) and experimental~\cite{Zipfel_PRB2022} (solid light) photoluminescence spectra (arbitrary units and vertical offset) for the attractive branch at different lattice temperatures. The theoretical spectra were obtained by convolving  Eq.~\eqref{eq:PLtotal} with a Lorentzian of width $1$~meV~\cite{Zipfel_PRB2022} and using $\varepsilon_T=22.5$ meV,     $n_h=0.5\times10^{11}$~cm$^{-2}$, and the MoSe$_2$ values of the exciton and hole effective masses: $m_X=1.15m_0$ and $m=0.59m_0$, with $m_0$ the free electron mass~\cite{Kormanyos_2DMat_2015}. 
    The experimental PL has been shifted horizontally to match the peaks of the virial expansion.}
    \label{fig:experiment_compare}
\end{figure}

\textit{Comparison with experiment}.---Recently, the PL originating from a MoSe$_2$ monolayer was measured for the case of a hole doping (per valley) of $n_h\simeq0.5\times10^{11}$~cm$^{-2}$ and for lattice temperatures $T=5$--$50$K~\cite{Zipfel_PRB2022}, corresponding to fugacities in the range $z\simeq 1$--$0.1$. Therefore, apart from the very lowest temperatures explored, the experiment was well within the regime of validity of the virial expansion. Furthermore we expect the role of phonons to be insignificant in this temperature range \cite{selig2016excitonic}, while Wigner crystallization is predicted to occur only below $10$K at this density~\cite{Zarenia2017}, in agreement with recent experimental observations~\cite{Smolenski_Nature2021}.  Therefore we only compare with experiment for $T\geq10$K. To compare our spectra calculated using Eq.~\eqref{eq:PLtotal}, we apply a Lorentzian broadening of 1~meV, matching the experimental linewidth~\cite{Zipfel_PRB2022}. 

We start by analyzing the distance between the peaks of the attractive and repulsive branches which, primarily due to the non-trivial shape of the attractive branch, does not correspond to $\varepsilon_T$ even at very low doping. Figure~\ref{fig:experiment_compare}(a) shows our theoretical result for two values of the trion binding energy, $\varepsilon_T=22.5$ meV and 23.5 meV, and for a range of densities. Even though this is noticeably below the quoted experimental value of 25 meV~\cite{Sidler_NatPhys_2017,Zipfel_PRB2022}, we see that the virial expansion correctly reproduces the splitting between the peaks when we take $\varepsilon_T=22.5$ meV. Thus, the fact that the attractive branch peak in PL does not correspond to the onset implies that the trion binding energy is likely to have been overestimated by as much as 10$\%$ in previous works~\footnote{This is also consistent with Ref.~\cite{Christopher2017} which has shown that the extracted trion binding energy is sensitive to the shape of the trion peak.}. We expect corrections to this result to be at most comparable to the Fermi energy~\cite{finiteTlong} which for this experiment is 0.4 meV.

Figure~\ref{fig:experiment_compare}(b) shows the comparison of our results for the attractive branch PL with experiment, using the extracted $\varepsilon_T$. We see that the agreement is essentially perfect at high temperature, with small discrepancies at lower temperatures. Since our theory is fully analytic and contains no free parameters, this is a remarkable agreement. The remaining discrepancy could potentially be due to the temperature of the system being different from that of the crystal lattice at low $T$.

{\it Connection to the trion picture.}---Our results for the attractive branch can be straightforwardly generalized 
beyond the low-energy expression in Eq.~\eqref{eq:2b_Tmatrix}.
In this case, we obtain $\mathcal{T}(\q_r;\q,\omega)$ from the spectral representation of the two-particle Green's function close to the trion pole, which finally gives~\cite{finiteTlong}
\begin{align}
    \Sigma_{\rm att}(\omega)&\simeq
    z\sum_\q e^{-\beta\epsilon_\q} |\eta_{\q_r}|^2 \nn \\
    & \hspace{-6.5mm}\times
    \left[{\mathcal P}\frac{(\epsilon_{r\q_r}+\varepsilon_T)^2}{\omega+\epsilon_{r\q_r}+\varepsilon_T}-i\pi\omega^2 \delta(\omega+\epsilon_{r\q_r}+\varepsilon_T)\right]\;,
    \label{eq:SigmaTrionWF}
\end{align}
where ${\mathcal P}$ denotes the principal value, $\eta_{\q_r}$ is the trion wave function, and $\epsilon_{r\q_r}=q_r^2/2m_r$. In general, 
we have $\eta_{\q_r}=\sqrt{Z_T(\q_r)}/(\varepsilon_T+\epsilon_{r\q_r})$~\cite{finiteTlong}, which yields Eq.~\eqref{eq:SEatt} in the low-energy limit where $Z_T$ can be approximated as the constant in Eq.~\eqref{eq:2b_Tmatrixpole}. 
Thus, to 
obtain the PL 
beyond the low-energy limit requires a knowledge of the trion wave function (a similar approach has been used to calculate absorption~\cite{Bronold2000}), which in general involves taking the details of the dielectric environment of the monolayer into account.

It turns out that previous trion theories of PL~\cite{Esser_PRB2000,Esser2001,Zipfel_PRB2022} correspond to the weakly interacting limit of our theory. Here, one assumes that the self-energy is sufficiently small such that the Dyson equation~\eqref{eq:dyson} can be expanded as $G_X(\omega)\simeq 1/\omega+\Sigma(\omega)/\omega^2$, which gives 
\begin{align} \nn
    P_{\rm att}(\omega)%
    \simeq z\, e^{\tfrac{\beta(m\omega -m_T\varepsilon_T)}{m_X}}\left|\eta_{\sqrt{2m_r|\omega+\varepsilon_T|}}\right|^2\Theta(-\omega-\varepsilon_T)
    \;,
\end{align}
where in the last step we used Eq.~\eqref{eq:SigmaTrionWF}. This precisely matches the result of Refs.~\cite{Esser_PRB2000,Esser2001}. Given the connection between the virial expansion and Fermi-polaron theory, we conclude that the trion picture results corresponds to a weak-coupling and high-temperature/low-doping limit of the Fermi-polaron picture, thus providing a formal unification of these two apparently disparate frameworks. Note that the weak-coupling assumption explicitly fails at the onset where the real part of the self-energy in Eq.~\eqref{eq:SigmaTrionWF} diverges, and hence the trion picture only  correctly describes the shape of the attractive branch in the limit $E_F \to 0$. Likewise, the broadening of the repulsive branch depends on exciton-electron scattering states which are neglected within trion-based theories~\cite{Bronold2000,Esser2001}.

\textit{Concluding remarks}.---
In summary, we have presented a controlled virial expansion for the exciton-polaron problem, which we show corresponds to a thermally incoherent limit of Fermi-polaron theory where the attractive polaron quasiparticle no longer exists. Our theory has the advantage of being fully analytic, and it yields excellent agreement with experiment without the need for fitting parameters. Our approach is very general and can be adapted to model the case of composite excitons with internal structure by using a modified trion wave function in Eq.~\eqref{eq:SigmaTrionWF}. Thus, it is potentially applicable to a broad range of systems, for instance emerging designer materials such as moir\'e superlattices where signatures of polaron physics have already been observed~\cite{Tang2020,Shimazaki2020,Campbell2022}.

\begin{acknowledgments} 
We are grateful to Dmitry Efimkin for useful discussions and feedback on our manuscript, and we thank Alexey Chernikov for sharing the data of Ref.~\cite{Zipfel_PRB2022}. 
BCM, MMP, and JL acknowledge support from the Australian Research Council Centre of Excellence in Future Low-Energy Electronics Technologies (CE170100039).
MMP and JL are also supported through the Australian Research Council Future Fellowships FT200100619 and FT160100244, respectively.
AT and FMM acknowledge financial support from the Ministerio de Ciencia e Innovaci\'on (MICINN), project No.~AEI/10.13039/501100011033 (2DEnLight).
FMM acknowledges financial support from the Proyecto Sinérgico CAM 2020 Y2020/TCS-6545 (NanoQuCo-CM).
\end{acknowledgments} 

%

\end{document}